\documentclass[aps,twocolumn,groupedaddress]{revtex4-1}

\usepackage{amsmath,amssymb,bm}
\usepackage{graphicx}
\usepackage[caption=false]{subfig}
\usepackage{bookman}
\usepackage{times}
\usepackage{ctable}
\usepackage{multirow}
\usepackage{floatrow}
\usepackage{wrapfig}
\usepackage{hyperref}

\begin{document}

\title{Time-dependent transport in open systems based on quantum master equations}

%\subtitle{Do you have a subtitle?\\ If so, write it here}

 % if too long for running head

\author{I. Knezevic}\email{knezevic@engr.wisc.edu}
\affiliation{Department of Electrical and Computer Engineering, University of Wisconsin - Madison, Madison, WI 53706, USA}
\author{B. Novakovic}
\affiliation{Department of Electrical and Computer Engineering, Purdue University, West Lafayette, IN 47907, USA} %etc.

\begin{abstract}
Electrons in the active region of a nanostructure constitute an open many-body quantum system, interacting with contacts, phonons, and photons. We review the basic premises of the open system theory, focusing on the common approximations that lead to Markovian and non-Mar\-ko\-vian master equations for the reduced statistical operator. We highlight recent progress on the use of master equations in quantum transport, and discuss the limitations and potential new directions of this approach.
\keywords{quantum transport \and master equation \and non-Markovian \and open system}
% \PACS{PACS code1 \and PACS code2 \and more}
% \subclass{MSC code1 \and MSC code2 \and more}
\end{abstract}

\maketitle
%%%%%%%%%%%%%%%%%%%%%%%%%%%%%%%%%%%%%%%%%%%%%%%%%%%%%%%%%%%%%%%%%%%%%
%%%%%%%%%%%%%%%%%%%%%%%%%%%%%%%%%%%%%%%%%%%%%%%%%%%%%%%%%%%%%%%%%%%%%
\section{Introduction}\label{sec:Introduction}
%%%%%%%%%%%%%%%%%%%%%%%%%%%%%%%%%%%%%%%%%%%%%%%%%%%%%%%%%%%%%%%%%%%%%
%%%%%%%%%%%%%%%%%%%%%%%%%%%%%%%%%%%%%%%%%%%%%%%%%%%%%%%%%%%%%%%%%%%%%

The term \textit{master equation} traditionally refers to differential equations that describe the time evolution of the probability that a given physical system will occupy a set of allowed states. These equations are typically first order in time and often, but not necessarily, linear in the probabilities. Nowadays, the term master equation is used more broadly: in the theory of few-level open systems, it refers to equations that describe the evolution of the open system's statistical operator (usually called the \textit{reduced statistical operator} or the \textit{reduced density matrix}) in the presence of coupling with an environment that has a large, perhaps infinite, number of degrees of freedom and is responsible for the irreversibility in the open system's evolution \cite{Alicki87,Breuer02}. Master equations are sometimes obtained as an ``educated guess" (i.e. phenomenologically), but can often be derived from the general framework of the open system evolution and a few reasonable assumptions. Master equations can be divided into \textit{Markovian master equations}, in which temporal evolution of the reduced statistical operator does not depend on its past but only its current state, and \textit{non-Markovian equations}, in which the so-called memory effects play an important role and involve information about the evolution of the environment.
How to best quantify non-Markovian effects in open quantum systems \cite{BreuerPRL_Non-Markovianity,PRL_LocalNonlocal,RivasPRL,CiracPRLForgetful} and how to experimentally control the information flow between the system and the environment, potentially driving the system between the Markovian to the non-Markovian regime \cite{NaturePhysicsNonMarkovianOptical}, are currently very active areas of inquiry.

Electronic systems in semiconductor nanostructures are open quantum systems, exchanging particles and information with the rapidly dephasing reservoirs of charge, often referred to as contacts, and possibly interacting with phonons or photons as well \cite{Frensley90}. In the open system theory, environments are commonly considered to be bosonic, which is fine for the interaction of electrons with light or lattice vibrations. However, electronic transport in the presence of contacts is a case of a fermionic open system coupled to fermionic reservoirs, which is a largely unexplored problem \cite{HadegardCaldeira,ShiArxiv}. When referring to electronic transport calculations, the use of the term master equation falls in two camps: on the one hand, we have few-level models (e.g. the resonant-level models used for quantum dots \cite{Gurvitz96}) for which master equations continue to refer to the dynamics of the reduced statistical operator. On the other hand, if we strive to account for the generally continuous single-particle energy spectrum of an electron in a nanostructure (e.g. when capturing current in structures without resonances) and the fact that many electrons are available to populate them, then calculating the full many-body reduced statistical operator becomes both intractable and unnecessary, as a great deal of information can be obtained from the single-particle quantities. In this case, master equation can refer to the equations for the time evolution of the single-particle density matrix (e.g. Redfield-type equations \cite{Ovchinikov05}) or just its diagonal terms (e.g. the Pauli master equation \cite{Fischetti99}).

In this paper, we review the basic premises and recent progress on the use of Markovian and non-Markovian master equations in the description of quantum electronic transport. In Sec. \ref{sec:general open sys formalism}, we present the basics of the open system formalism, including the concept of complete positivity of a dynamical map. We discuss microscopic derivations of Markovian master equations in Sec. \ref{sec:general Markovian}, focusing on the weak-coupling limit, and follow with examples from quantum transport in Sec. \ref{sec:examples Markovian}. In Sec. \ref{sec:general non-Markovian}, we overview general features of non-Markovian master equations and present the Nakajima-Zwanzig projection operator technique. Examples of non-Markovian master equations for time-dependent quantum transport are given in Sec. \ref{sec:examples non-Markvian}. A summary and outlook conclude this paper in Sec. \ref{sec:Conclusion}.

\section{General open system formalism}\label{sec:general open sys formalism}

Consider a quantum-mechanical system $S$ interacting with an environment $E$. This composite $S+E$ system is generally described by the full statistical operator that, like other operators, lives in the Liouville space, which is (in the case of finite systems) isomorphic to the square of the composite Hilbert space, ${\mathcal{H}}^2=(\mathcal{H}_S\otimes\mathcal{H}_E)^2$. Here, $\mathcal{H}_S$ and $\mathcal{H}_E$  are the Hilbert spaces while $\mathcal{H}_S^2$ and $\mathcal{H}_E^2$ are the Liouville spaces of the system and environment, respectively. Operators acting on the Liouville space are often called \textit{superoperators}. If $S+E$ is closed, the dynamics of its statistical operator $\rho$ is given by the Liouville equation (in the units of $\hbar=1$)
\begin{equation}\label{eq:Liouville S+E}
\frac{d\rho}{dt}=-i\mathcal{L}\rho=-i\left[H,\rho\right].
\end{equation}
$\mathcal{L}$ is the Liouville superoperator and $H$ the total $S+E$ Hamiltonian, generally of the form $H=H_{S}\otimes I_E+I_{S}\otimes H_E+H_{\mathrm{int}}$. The integral form of the Liouville equation is
\begin{subequations}
\begin{eqnarray}\label{eq:unitary}
\rho(t)&=&U(t,t_0)\rho(t_0)U^\dagger(t,t_0),\\
U(t,t_0)&=&\mathrm{T}\exp{\left(\int_{t_0}^{t}-iH(t) dt\right)},
\end{eqnarray}
\end{subequations}
with $\mathrm{T}$ denoting time ordering. In the case of a time-in\-de\-pen\-dent $H$, $U(t,t_0)=\exp\left(-iH(t-t_0)\right)$.

Assume that we are interested only in the evolution of $S$. Its statistical operator can be obtained by tracing out the $E$ degrees of freedom, i.e.,
\begin{equation}\label{eq:trace over E}
\rho_S=\mathrm{Tr}_E\left(\rho\right).
\end{equation}
$S$ is often referred to as the \textit{reduced system} and $\rho_S$ as the \textit{reduced statistical operator} or the \textit{reduced density matrix}. We will use the term reduced statistical operator, because
the term density matrix is usually reserved for the single-particle quantity in quantum transport studies.

While the dynamics of $S+E$, a closed system, is unitary, the dynamics of $S$ is not. If the environment has a large number of degrees of freedom, on the timescales accessible in experiment the evolution of the reduced system $S$ effectively becomes irreversible. Quite generally, the dynamics of the reduced statistical operator is given by
\begin{subequations}
\begin{eqnarray}
\rho_S(t)&=&\mathrm{Tr}_E\left(U(t,t_0)\rho(t_0)U^\dagger(t,t_0)\right)\\
\frac{d\rho_{S}}{dt}&=&-i\mathrm{Tr}_E\mathcal{L}\rho=-i\mathrm{Tr}_E\left[H,\rho\right].
\end{eqnarray}
\end{subequations}
The central goal of the open system theory is to obtain the evolution of the (relatively small) reduced system $S$ while minimizing the information that has to be gathered about the (relatively large) environment $E$. This quest is understandably very difficult, and approximations must be employed to yield tractable equations.

A first major simplification is assuming that, at some point in the $S+E$ evolution, the $S$ and $E$ were decoupled.
It is assumed that, up until a certain point in time, usually designated as $t=0$, $S$ and $E$ were mutually isolated,
non-interacting, and therefore the initial $\rho$ is of the uncoupled, tensor-product form,
\begin{equation}\label{eq:uncopled initial state}
\rho(0)=\rho_{S}(0)\otimes\rho_{E}(0).
\end{equation}
\noindent Thereafter, the interaction is turned on, presumably adiabatically. (A reader interested in the field of nanoelectronics can immediately ask if this assumption is ever satisfied in electronic systems, and the answer is ``sometimes." For instance, it can be considered true when we have high tunnel barriers between the active region ($S$) and contacts ($E$) in a nanostructure, and have let the active region and environment each relax on its own, with minimal tunneling between them.)

The assumption of an uncorrelated initial state is a very appealing one to adopt, because it guarantees \cite{Lindblad96} the existence of a \textit{subdynamics} (also known as \textit{reduced dynamics}), i.e. it guarantees that the evolution of the reduced statistical operator can in principle be fully described within $\mathcal{H}_S^2$. In other words, the existence of a subdynamics means there exists a generally nonunitary evolution superoperator $\mathcal{W}(t,0)$, such that
\begin{equation}
\rho_S(t)=\mathcal{W}(t,0)\rho_S(0).
\end{equation}
$\mathcal{W}(t,0)$ is often referred to as a \textit{dynamical map}. In general, there exists a non-Hermitian \textit{generator of the dynamical map}, $\mathcal{K}(t)$, which satisfies
\begin{equation}
\mathcal{W}(t,0)=\mathrm{T}\int_0^t \mathcal{K}(t')\,dt';\,  \mathcal{K}(t)=\frac{d\mathcal{W}}{dt}.
\end{equation}

\noindent The differential equation of motion for a subdynamics can be written as
\begin{equation}\label{eq:subdynamics differential form}
\frac{d\rho_S}{dt}=\mathcal{K}(t)\rho_S(t).
\end{equation}

Clearly, it is practically impossible to obtain $\mathcal K$ and $\mathcal W$ from first principles, and approximations are commonly made to the structure of $\mathcal K$. Generator $\mathcal K$ should be such that the resulting evolution does not violate the unit trace or the positivity of $\rho_S$. While the former is quite easily satisfied (any generator that acts as a commutator or a sum of commutators will preserve the trace of $\rho_S$), the latter is generally a tall order and is, in fact, not fulfilled by many common approximations. \textit{Complete positivity} of a dynamical map \cite{Gorini76} is a stronger criterion than positivity (i.e. requiring that the map preserve the positivity of the statistical operator). Namely, if you have two systems whose evolution is such that the density matrix of each remains positive (i.e. the evolution operators $\mathcal W$ for each subsystem are positive maps), it is not guaranteed that the composite map (their tensor product) will be a positive map, i.e. it is not guaranteed that the composite statistical operator will remain positive throughout evolution. If, however, each one of the evolution maps is completely positive, then the tensor product is completely positive. In essence, complete positivity of the evolution map is a stricter criterion than positivity and necessary for a successful description of composite systems \cite{Alicki87}.

A time-independent generator $\mathcal K$ corresponds to Markovian approximations. Evolution operators $\mathcal W(t,t')$ generated by a time-independent $\mathcal K$ form a semigroup, with $\mathcal K$ then referred to as the semigroup generator. It has been shown by Lindblad \cite{Lindblad76} that the most general case of a generator of a completely positive Markovian evolution must be of the form (given in the Schr\"{o}dinger picture):
\begin{equation}\label{eq:Lindblad equation}
\mathcal{K}\rho_S=-i[H_S,\rho_S]+\sum_{k} \gamma_k \left([A_k, \rho_S A_k^\dagger]+[A_k\rho_S , A_k^\dagger]\right),
\end{equation}
\noindent where $\gamma_k$ are nonnegative coefficients. The last term on the right-hand side is often referred to as the \textit{dissipator}. The dynamical map $\mathcal{W}(t,t')$ generated by $\mathcal K$ from Eq. (\ref{eq:Lindblad equation}) is a completely positive Markovian map.
Lindblad's form of $\mathcal K$ is very useful because it enables development of physically reasonable approximate forms of semigroup generators.

\section{Microscopic derivations of Markovian master equations}\label{sec:general Markovian}

In the most general terms, completely positive Markovian equations for the subsystem dynamics can be obtained in the weak-coupling limit \cite{Davies74,VanHove55,VanHove59,Chester63}, singular coupling limit \cite{Breuer02}, and by coarse graining over time \cite{Lidar01} (discussed in more detail in Sec. \ref{sec:Coarse graining}). In electronic systems, there is also the large bias limit with contacts that have a constant density of states \cite{Gurvitz96} (also referred to as the \textit{wide-band limit} \cite{Haug98,StefanuccivanLeeuwen}),  which we will discuss separately. The weak-coupling limit is of particular importance, being applicable to electron-phonon interaction  and electron-contact coupling in the case of tunnel barriers, so we discuss it in more detail.

\subsection{The weak-coupling limit}

The total dynamics in the interaction picture can be written as (in differential and integral forms)
\begin{subequations}
\begin{eqnarray}
\frac{d}{dt}\rho(t)&=&-i[H_{I}(t), \rho(t)],\\
\rho(t)&=&\rho(0)-i\int_0^t ds\,[H_I(s),\rho(s)],
\end{eqnarray}
\end{subequations}
where $H_I(t)$ is the interaction Hamiltonian in the interaction picture. Putting the integral form in the right-hand side of the differential form results in $\frac{d\rho(t)}{dt}=-i[H_{I}(t), \rho(0)]-\int_0^t ds\,[H_I(t),[H_I(s),\rho(s)]$. Tracing out this equation over the environment degrees of freedom, we obtain
\begin{eqnarray}
\frac{d\rho_{S}(t)}{dt}&=&-i\mathrm{Tr}_E[H_{I}(t), \rho(0)]\\ \nonumber
&-&\int_0^t ds\, \mathrm{Tr}_E[H_{I}(t),[H_I(s),\rho(s)]].
\end{eqnarray}

\noindent It is commonly assumed that $\mathrm{Tr}_E[H_{I}(t), \rho(0)]=0$. This assumption is often satisfied: for instance, if the initial statistical operators of the system and environment are grand-canonical or canonical equilibrium ensembles, they will contain pairs of the creation and annihilation operators associated with the single-particle spectra for $S$ and $E$. The interaction Hamiltonian is usually linear in these operators, i.e. it is  commonly assumed to be of the form (in the Schr\"{o}dinger picture) $\sum_{\alpha}A_{\alpha}\otimes B_{\alpha}$, where $A_\alpha$ are the system and $B_\alpha$ the environment operators. Tracing out the product of this interaction Hamiltonian with the environment statistical operator over environmental states gives zero.

In electronic systems, the above approximation is satisfied for the electron-phonon interaction (the interaction Ha\-mil\-to\-nian is linear in phonon creation and annihilation operators) as well as for typical model Hamiltonians that describe the interaction of the device with the contacts (Hamiltonian linear in the contact and device creation/annihilation operators) if the evolution starts from equilibrium. However, if contact-active region electron-electron interaction is deemed important and is part of the interaction Hamiltonian, then the term $\mathrm{Tr}_E[H_{I}(t), \rho(0)]$ would survive.

So the equation we are focusing on, which is still exact provided approximation $\mathrm{Tr}_E[H_{I}(t), \rho(0)]=0$ holds, is:
\begin{equation}
\frac{d\rho_{S}(t)}{dt}=-\int_0^t ds\, \mathrm{Tr}_E[H_{I}(t),[H_I(s),\rho(s)]].
\end{equation}

The \textit{Born approximation} assumes that the interaction is weak, so that the environment is negligibly  affected by it and no considerable $S$-$E$
correlations arise due to it over time on the timescales relevant to $S$. As a result, we can write $\rho(t)\approx \rho_{S}(t)\otimes\rho_E(0)$ and, consequently,

\begin{equation}\label{eq:Born only}
\frac{d\rho_{S}(t)}{dt}=-\int_0^t ds\, \mathrm{Tr}_E[H_{I}(t),[H_I(s),\rho_S(s)\otimes\rho_E]]
\end{equation}

\noindent Equation (\ref{eq:Born only}) has memory. The \textit{Markov approximation} assumes that the interaction magnitude is such that the evolution will depend only on the present state of the system, not its prior evolution, so $\rho_S(s)$ is replaced by $\rho_S(t)$. This assumption is valid on timescales coarser than the decay time of environmental correlations \cite{Breuer02}. We can switch from $s$ to $t-s$, with $s$ now denoting the temporal distance from $t$, and integrate over all values of $s$, because we expect the integrand to be negligible for large values of $s$ (i.e. environmental correlations decay rapidly with increasing $s$), finally arriving at
\begin{equation}\label{eq:Redfield}
\frac{d\rho_{S}}{dt}=-\int_0^\infty ds\, \mathrm{Tr}_E[H_{I}(t),[H_I(t-s),\rho_S(t)\otimes\rho_E]].
\end{equation}
\noindent Equation (\ref{eq:Redfield}) is the \textit{Redfield equation} \cite{Redfield57} and it still has memory.

For an interaction Hamiltonian of the form $\sum_{\alpha}A_\alpha\otimes B_\alpha$, we can define the Fourier transforms of $A_\alpha$ and $B_\alpha$ based on the system and environment spectra,
\begin{equation}
A_{\alpha}(\omega)=\sum_{\varepsilon}\mathcal P (\varepsilon)A_{\alpha}\mathcal P (\varepsilon +\omega),
\end{equation}
where $\mathcal P (\varepsilon)$ projects onto the eigenspace of $H_S$ corresponding to eigenvalue $\varepsilon$. As a result, $A_{\alpha}(t)=e^{iH_S t}A_{\alpha}(\omega)e^{-iH_S t}\\=e^{-i\omega t}A_{\alpha}(\omega)$. In that case, the interaction Hamiltonian in the interaction picture becomes $H_I(t)=\sum_{\alpha,\omega}e^{-i\omega t}A_{\alpha}(\omega)\otimes B_{\alpha} (t)$, where $B_{\alpha} (t)=e^{iH_E t}B_{\alpha}e^{-iH_E t}$. Finally, the evolution of the reduced statistical operator becomes

\begin{eqnarray}
\frac{d}{dt}\rho_S(t)&=&\sum_{\omega ,\omega'}\sum_{\alpha ,\beta}e^{i(\omega -\omega')t}
\Gamma_{\alpha\beta}(\omega)\left(A_\beta (\omega)\rho_S(t)A^{\dagger}_\alpha(\omega')\right.\nonumber\\
&-& \left. A^{\dagger}_\alpha(\omega')A_\beta (\omega)\rho_S(t)\right)+\mathrm{h.c.}\, ,
\end{eqnarray}

\noindent where $\Gamma_{\alpha\beta}(\omega)=\int_0^\infty ds\, e^{i\omega s}\,\mathrm{Tr}_E\left(B^\dagger_\alpha (t)B_{\beta}(t-s)\right)$. When the typical timescales for the system evolution, proportional to $|\omega -\omega '|^{-1}$, are much shorter than the expected relaxation timescales for the system, the so-called\textit{ secular approximation }(also known as the \textit{rotating-wave approximation} or RWA) can be applied: all terms with $\omega -\omega ' \neq 0$ are considered as varying too fast, so that their average contribution on the timescales relevant to $S$ can be neglected. As a result, we obtain the weak-limit Markovian equation of motion
\begin{eqnarray}\label{eq:Markovian weak coupling after RWA}
\frac{d}{dt}\rho_S(t)&=&\sum_{\omega }\sum_{\alpha ,\beta}
\Gamma_{\alpha\beta}(\omega)\left(A_\beta (\omega)\rho_S(t)A^{\dagger}_\alpha(\omega)\right.\nonumber\\
&-& \left. A^{\dagger}_\alpha(\omega)A_\beta (\omega)\rho_S(t)\right)+\mathrm{h.c.}\, .
\end{eqnarray}

\noindent Let us define
\begin{subequations}
\begin{eqnarray}
\chi_{\alpha\beta}(\omega)&=&\frac{1}{2i}\left(\Gamma_{\alpha\beta}(\omega)-\Gamma^{*}_{\beta\alpha}(\omega)\right)\\
\gamma_{\alpha\beta}(\omega)&=&\frac{1}{2}\left(\Gamma_{\alpha\beta}(\omega)+\Gamma^{*}_{\beta\alpha}(\omega)\right).
\end{eqnarray}
\end{subequations}

\noindent $\chi$ corresponds to the so-called \textit{Lamb shift}, an effective correction to the system Hamiltonian of the form
\begin{equation}
H_{LS}=\sum_{\omega ,\alpha ,\beta}\chi_{\alpha\beta}(\omega)A^\dagger_{\alpha}(\omega)A_{\beta}(\omega)\, .
\end{equation}
$H_{LS}$ commutes with $H_S$, so it shares the eigenvectors with $H_S$ and simply corrects the $H_S$ energy levels, and is therefore not a true dissipative term.  $\gamma$ defines the coefficient of the true dissipator,
\begin{eqnarray}\label{eq:Lindblad dissipator}
\mathcal{D}(\rho_S)&=&\sum_{\omega}\sum_{\alpha ,\beta}
\gamma_{\alpha\beta}(\omega)\left([A_\beta (\omega)\rho_S(t),A^{\dagger}_\alpha(\omega)]\right.\nonumber\\
&+&\left.[A_\beta (\omega),\rho_S(t)A^{\dagger}_\alpha(\omega)]\right).
\end{eqnarray}

\noindent This dissipator is of Lindblad form, which can be shown after proving that $\gamma_{\alpha\beta}$ is positive definite and diagonalizing it.

If the system Hamiltonian is diagonalized in a basis $| n\rangle$ as $H_S=\sum_{n}\varepsilon_n | n\rangle \langle n |$, then we can derive an equation of motion for the populations of the eigenstates $\rho_S(n,t)=\langle n |\rho_S (t)|n\rangle$ as

\begin{equation}\label{eq:Pauli}
\frac{d\rho_S(n,t)}{dt}=\sum_{n'}S(n,n')\rho_S(n,t)-S(n',n)\rho_S(n',t),
\end{equation}

\noindent $S(n,n')=\sum_{\alpha\beta}\gamma_{\alpha}(\varepsilon_n' - \varepsilon_n)\langle n' | A_{\alpha} |n\rangle \langle n | A_{\beta} |n'\rangle$ being the transition rates obtained from Fermi's golden rule \cite{Breuer02}. Equation (\ref{eq:Pauli}) is known as the \textit{Pauli master equation}.

\section{Examples of Markovian master equations in quantum transport}\label{sec:examples Markovian}

\subsection{Pauli master equation for electron-phonon interaction}\label{sec:Pauli}

An example of the Pauli master equation in the treatment of electron-phonon interaction in devices is the work of Fischetti \cite{Fischetti98,Fischetti99}.
He has shown that, in the Born-Markov approximation and the van Hove limit (time tends to infinity while the coupling strength tends to zero,
so that interaction squared times time remains constant and nonzero during the limiting procedure)\cite{Davies74}, the master equation for the fermionic active region will include the
exclusion principle, thus generally becoming non-linear for high population of states. Scattering states $|\mu\rangle$ that diagonalize the single-electron Hamiltonian
in the active region can be obtained from the solution of the coupled Schr\"{o}dinger and Poisson equations with open boundary conditions. In order to accurately compute spatially resolved quantities,
such as charge density and potential, in the numerical implementation, an appropriately dense set of scattering states is obtained through a mapping onto standing-wave-type solutions (details can be found in \cite{Fischetti98}).

In the work, the active region -- contact interaction is treated through a boundary injection/collection term that acts as a source to the equation.
The Pauli master equation in the basis of scattering states reads

\begin{eqnarray}\label{eq:Pauli Fischetti}
&\frac{\partial \rho_S(\mu,t)}{\partial t}=\sum_{\lambda}S(\mu,\lambda)\rho_S(\mu,t)[1-\rho_S(\lambda,t)]\\
&-S(\lambda ,\mu)\rho_S(\lambda,t)[1-\rho_S(\mu,t)]+\left(\frac{\partial \rho_S(\mu ,t)}{\partial t}\right)_{\mathrm{con.}},\nonumber
\end{eqnarray}

\noindent where the source term for contact $j$ is given by

\begin{equation}
\left(\frac{\partial \rho_S^{(j)}(\mu ,t)}{\partial t}\right)_{\mathrm{con.}}\sim \upsilon_{\perp}(\vec{k}_{\mu ,j})[f^{(j)}(\vec{k}_{\mu ,j})-\rho_S^{(j)}(\mu ,t)].
\end{equation}

\begin{figure}
\includegraphics[width=3 in]{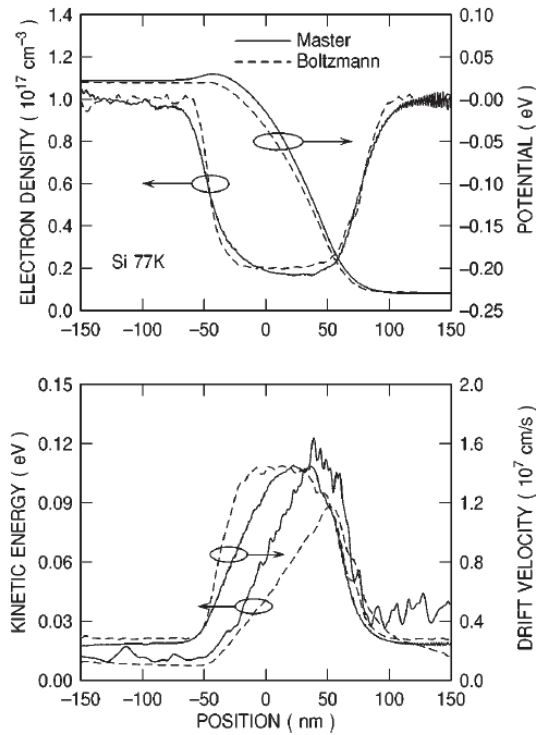}
\caption{Calculated potential energy, electron charge density (top panel), average drift velocity  and average kinetic energy (bottom panel) for an \textit{nin} silicon diode at 77 K biased to 0.25 V. The solid lines refer to results calculated using the master equation, the dashed lines to results obtained using a semiclassical full-band Monte Carlo simulation employing identical parameters. Reprinted with permission from Ref. \cite{Fischetti99},
M. V. Fischetti, Phys. Rev. B 59, 4901 (1999). (c) 1999 The American Physical Society}\label{fig:TwoTerminalSchematic}
\end{figure}

\noindent Here, $f^{(j)}(\vec{k}_{\mu ,j})$ is the distribution function in contact $j$ and $\upsilon_{\perp}(\vec{k}_{\mu ,j})$
is the perpendicular component of velocity associated with state $\mu$ and normal to the active region/contact $j$ boundary.
Figure \ref{fig:TwoTerminalSchematic} shows a comparison between the Pauli master equation and ensemble Monte Carlo simulation of a silicon \textit{nin} diode.

There is a concern that the Pauli master equation does not conserve current outside of the steady state.
It has been shown that current is conserved as long as coupling to the contacts is local \cite{Gebauer04}.

\subsection{Markovian equations for system-environment coupling}\label{sec:examples Markovian for system environment}

One of the early contributions aimed specifically at the treatment of transport in electronic systems via master equations was the paper by Gurvitz and Prager \cite{Gurvitz96}.
In their work, the approximation of high bias has enabled the Markov approximation. They discuss resonant transfer in mesoscopic devices,
focusing on resonant states as the only relevant eigenstates of the electronic Hamiltonian in the systems of interest.
The resonant level model is commonly adopted \cite{Wingreen93,Jauho94,Haug98}. The open system has two terminals and is coupled
only to the left and right reservoirs, such that the resonant levels are comfortably inside the transport window
(the range of energies between the Fermi levels of the two contacts) and the density of contact states is constant throughout.
Markovian evolution can be obtained in the form of the density matrix in the basis of the resonant states,
with off-diagonal terms making it different from the phenomenological rate equations \cite{Nazarov93}. Around the same time, Stoof and Nazarov \cite{StoofNazarov96} investigated time-dependent resonant tunneling via two discrete states in the presence of resonant-frequency irradiation based on a phenomenological Markovian master equation for the full statistical operator of this two-level system.

An enhancement to the work of Gurvitz and Prager \cite{Gurvitz96} was put forth by Li \textit{et al.} \cite{Li05} for a system with multiple resonances,
such as a quantum dot, that is connected to the reservoirs via barriers through which tunneling is relatively weak.
Starting from the Born approximation and working with conditional density matrices that correspond to a fixed number of electrons getting onto the dot at a given time,
the authors are able to derive a Markovian equation of the Lindblad form that does not require the wide-band limit \cite{Gurvitz96}.

The work of Harbola, Esposito, and Mukamel \cite{Harbola06} uses projection operators (see more in Sec. \ref{sec:Nakajima Zwanzig} below)
to derive a hierarchy of quantum master equations for the many-body statistical operators
representing the system with a given number of electrons. They show that Fock-space coherences between states with different populations do not contribute to transport
to second order in system-environment coupling, but coherences between different many-body states with the same \textit{n} are appreciable.

Espostio and Galperin \cite{Esposito10} derived a time-local Markovian master equation for molecular transport based on the Redfield equation,
which is nonlocal in time, and supplanting it with a kind of time-reversed Redfield evolution that enables a self-consistent procedure for deriving the generator.

\begin{figure}
\includegraphics[width=3 in]{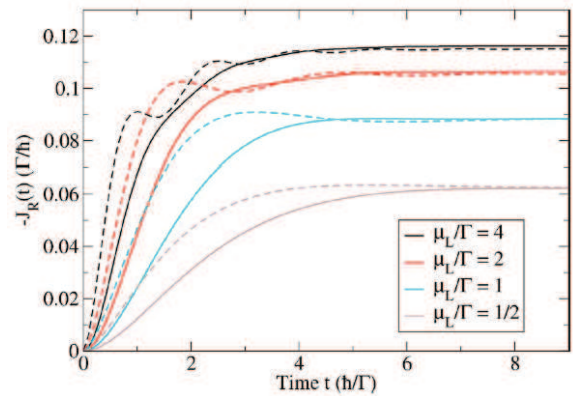}
\caption{The time-dependent current calculated with the method of Pedersen and Wacker \cite{Pedersen05} (solid line) and with the time-dependent Green function method \cite{StefanucciTDDFTPRB04} (dashed line) in response to
steplike modulation of the bias, with step height
$\mu_L$. The coupling to the left and right contacts are $\Gamma_L=\Gamma_R=\Gamma/2$,
the temperature is $k_BT=0.05\Gamma$, and the half-width of the
band is $W=30\Gamma$. Reprinted with permission from Ref. \cite{Pedersen05}, J. N. Pedersen and A. Wacker, Phys. Rev. B 72,  195330 (2005). (c) 2005 The American Physical Society}\label{fig:PedersenWacker}
\end{figure}

Pedersen and Wacker \cite{Pedersen05} worked in the basis of the full many-body Hamiltonian and derived Markovian master equations for few-level systems coupled to a continuum of lead states.
The long-time evolution coincides with the non-Markovian description based on time-dependent Green's functions (see Fig. \ref{fig:PedersenWacker}).
The evolution they describe is numerically tractable and contains considerably more information than the rate equations. The approach is referred to as the second-order von Neumann approach (2vN for short), indicating that the correlations between two tunneling events are included \cite{WackerPhysicaE2010}. Based on a diagrammatic expansion, Karlstr\"{o}m \textit{et al.} recently showed the equivalence between the 2vN approach and the resonant tunneling approximation, and discussed the limitations of the technique in the calculation of higher order cumulants \cite{Wacker2013}.

\section{Micoscopic derivations of non-Markovian equations}\label{sec:general non-Markovian}

It is known that the general, completely positive, non-Mar\-ko\-vian evolution of an open system that started in an uncorrelated state (\ref{eq:uncopled initial state})
can be written as
\begin{equation}
\rho_S(t)=\sum_i R_i(t)\rho_S(0)R_i^\dagger (t).
\end{equation}
This form is usually referred to as the \textit{operator-sum representation} or \textit{Kraus representation} \cite{Kraus71}, where $R_i(t)$ are the Kraus operators.
Approximate Kraus maps based on physically reasonable assumptions have been constructed \cite{VanWonderen05}.
However, in contrast to Markovian evolution, where the Lindblad equation (\ref{eq:Lindblad equation})
specifies the required form for a generator of a completely positive dynamical map, there are no similarly
compact criteria to determine if an approximate non-Markovian map is completely positive or not.

\subsection{Nakajima-Zwanzig and time-convolutionless (TCL) projection operator techniques}\label{sec:Nakajima Zwanzig}

A general and widely applied technique for the derivation of non-Markovian master equations up to a given order in the $S-E$ interaction is the Nakajima-Zwanzig projection operator technique \cite{Nakajima58,Zwanzig60}. Commonly, terms up to the second or fourth order in the interaction are retained, but complete positivity of the resulting master equations is generally not guaranteed.

In the $S+E$ Liouville space $\mathcal{H}^2$, any environment density $\rho_E$ matrix generates a projection operator $\mathcal P$ whose action is given by $\mathcal{P}x=\mathrm{Tr}_E(x)\otimes\rho_E
$, $x\in \mathcal{H}^2$. ${\mathcal P}$ is a projector, meaning that ${\mathcal P}^2={\mathcal P}$. The range (space of images) of ${\mathcal P}$  is isomorphic to the system
Liouville space $\mathcal{H}_S^2$. The complementary projector is $\mathcal Q =1-\mathcal P$.

By projecting the Liouville equation (\ref{eq:Liouville S+E}) onto the ranges of $\mathcal P$ and $\mathcal Q$, we obtain two equations of motion
\begin{subequations}
\begin{eqnarray}
i\frac{\partial}{\partial t}\mathcal P \rho=\mathcal P\mathcal L \mathcal P\rho+\mathcal P\mathcal L\mathcal Q\rho ,\label{eq:projection op 1st eq}\\
i\frac{\partial}{\partial t}\mathcal Q \rho=\mathcal Q\mathcal L\mathcal P\rho+\mathcal Q\mathcal L\mathcal Q\rho .
\end{eqnarray}
\end{subequations}

\noindent If the interaction Hamiltonian, in the interaction picture, is of the form $\epsilon H_I(t)$, where $\epsilon $ is a unitless number characterizing the smallness of the interaction,
we can formally solve the equation for $Q\rho$ as
\begin{equation}
\mathcal{Q}\rho_S(t)=\mathcal{G}(t,t_0)\mathcal Q\rho_S(t)+\epsilon\int_{t_0}^t ds\, \mathcal{G}(t,s)\mathcal Q\mathcal L (s)\mathcal P\rho_S(s),
\end{equation}

\noindent where $\mathcal G (t,s)=\mathrm{T}\exp{\left[\epsilon\int_s^t ds'\,\mathcal Q\mathcal L (s')\right]}$. Substituting this equation into
(\ref{eq:projection op 1st eq}) above, we obtain the \textit{Nakajima-Zwanzig equation}:

\begin{eqnarray}
\frac{d}{dt}\mathcal{P}\rho_S&=&\epsilon\mathcal{P}\mathcal{L}(t)\mathcal{G}(t,t_0)\mathcal{Q}\rho(t_0)+\epsilon\mathcal{P}\mathcal{L}(t)\mathcal{P}\rho_S(t)\nonumber\\
&+&\epsilon^2\int_{t_0}^t ds\, \mathcal{P}\mathcal{L}(t)\mathcal{G}(t,s)\mathcal{Q}\mathcal{L}(s)\mathcal{P}\rho_S(s) .
\end{eqnarray}

\noindent Commonly, in the case of an uncorrelated initial state (\ref{eq:uncopled initial state}), the initial environment density matrix is chosen to generate the
projection operator $\mathcal P$, which means that $\mathcal{P}\rho(0)=\rho(0)$ and $\mathcal{Q}\rho(0)=0$. Alternatively, the projector may be chosen so as to annul the odd moments of the interaction Hamiltonian.
The choice of $\mathcal{P}$ generally depends on the application in mind, and $\mathcal{P}$ is often assumed to be associated with the equilibrium canonical or grand canonical statistical operator for the environment.

The \textit{time-convolutionless (TCL) projection operator technique}, originally  due to Shibata \textit{et al.} \cite{Shibata77}, writes the Na\-ka\-ji\-ma-Zwanzig equation in a form that depends only on the instantaneous $\rho_S(t)$, and all the memory effects are relegated to certain evolution operators, which opens doors to systematic approximations, even if the operators are still quite unwieldy and a partial trace does technically need to be taken over the equation after everything. (Partial-trace-free time-convolutionless equations of motion and the related concept of memory dressing have been proposed in \cite{Knezevic02,Knezevic04}.)
Here, we quote the TCL equation in the form without the inhomogeneity, i.e. for $\mathcal{Q}\rho(0)=0$.
\begin{equation}
\frac{d}{dt}\mathcal{P}\rho(t)=\mathcal{K}(t)\mathcal{P}\rho(t),
\end{equation}
where
\begin{subequations}
\begin{eqnarray}
\mathcal{K}(t)&=&\epsilon\mathcal{P}\mathcal{L}(t)[1-\Sigma(t)]^{-1}\mathcal{P},\\
\Sigma(t)&=&\epsilon\int_{t_0}^{t} ds\, \mathcal G(t,s)\mathcal Q\mathcal L(s)\mathcal P \mathcal U(t,s),\\
\mathcal U(t,s)&=&\mathrm{T}\exp{\left[-\epsilon\int_s^t ds'\,\mathcal{L}(s)'\right]}.
\end{eqnarray}
\end{subequations}
Obviously, there is an assumption that $1-\Sigma(t)$ is invertible \cite{Knezevic02}. Upon performing a Taylor expansion of
$1-\Sigma(t)$ in terms of $\epsilon$, we can get a series $\mathcal{K}(t)=\sum_{n}\epsilon^n \mathcal{K}_{n}(t)$, where $\mathcal{K}_{1}=0$, $\mathcal{K}_{3}=0$,
$\mathcal{K}_{2}(t)=\int_{t_0}^t dt'\, \mathcal{P}\mathcal{L}(t)\mathcal{L}(t')\mathcal{P}$, and
$\mathcal{K}_{4}(t)=\int_{t_0}^t dt_1\int_{t_0}^t dt_2 \int_{t_0}^t dt_3\, \mathcal{P}\mathcal{L}(t)\mathcal{L}(t_1)\mathcal{L}(t_2)\mathcal{L}(t_3)\mathcal{P}$.

If $\mathcal{K}\approx\mathcal{K}_2$, the TCL equation yields the Redfield equation (\ref{eq:Redfield}). A number of examples of TCL equations with second
and fourth order coupling can be found in \cite{Breuer02}.  Timm \cite{Timm08} discusses a diagrammatic expansion of time-convolutionless equations.

It is also worth noting that a projection operator technique can be used to derive the well-known semiconductor Bloch equations \cite{LindbergKochSemiBlochEqns88}.

\section{Examples of Non-Markovian transport equations in quantum transport}\label{sec:examples non-Markvian}

One of the early non-Markovian approaches to electron transport in nanostructures was put forth by Bruder and Schoeller \cite{Bruder94}. Time-dependence was introduced either by periodic modulation of the Fermi energy
or by time-dependent perturbations to the quantum states in the dot. The authors focused on the effects of the Coulomb interaction
in the limit of low tunneling rates but finite level spacing.

\begin{figure}
\includegraphics[width=\columnwidth]{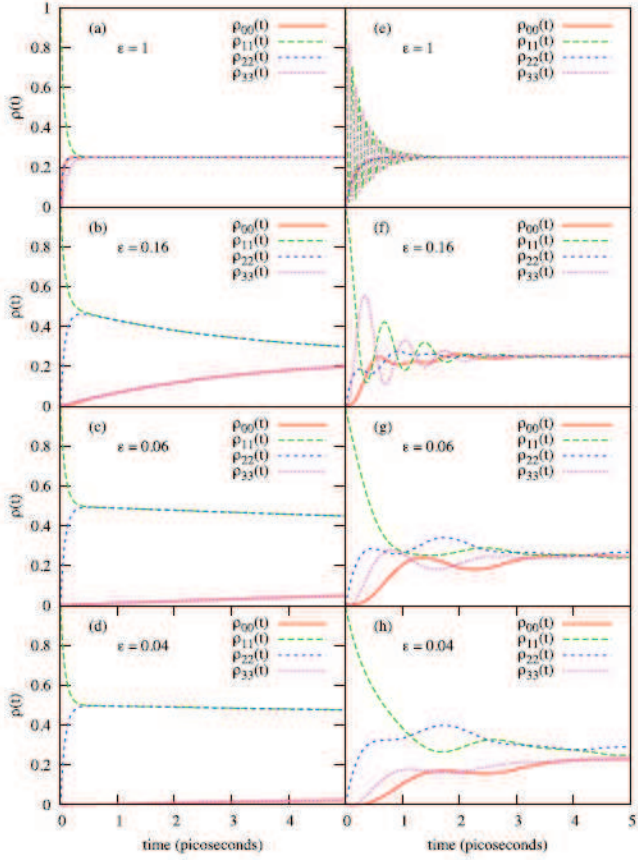}
\caption{Markovian and non-Markovian time
evolution of population probabilities in a quantum dot with two transport channels and four states. The plots
are for symmetric source and drain tunnel barriers, and varying orbital asymmetry. A 6 meV bias symmetric about the Fermi energy is assumed. The two transport channels have energies $\pm$1 meV around the Fermi level.
Plots (a) through (d) are results for the Markov limit, whereas plots (e) through (h) present results for the non-Markovian theory.
Reprinted with permission from \cite{Vaz10PRB}, E. Vaz and J. Kyriakidis, Phys. Rev. B 81, 085315 (2010). (c) 2010 The American Physical Society}\label{fig:Vaz}
\end{figure}

Vaz and Kyriakidis \cite{Vaz08JPCS,Vaz10PRB,Vaz2012PRB} calculated the full Redfield tensor in Fock space for a two-level system (Fig. \ref{fig:Vaz}).
The authors find that Fock-space coherences between states with different particle numbers are robust and may be preserved
even in the presence of tunneling into and out of the dot. The authors also note that, while Redfield dynamics could potentially violate positivity of
the statistical operator, they have not observed it in practice \cite{Vaz08JPCS}.

Recently, Gudmundsson and co-authors \cite{GudmundssonNJP09} used a non-Markovian transport equation to analyze time-dependent transport in a few-mode nanowire containing a localized region
and focused on the effect of nontrivial geometry. The authors pay attention to the fact that arbitrary decisions where the active region ends and contacts begin lead to inconsistencies,
and that an effective overlap between the $S$ and $E$ wave functions will yield effective interaction Hamiltonian matrix elements. This important issue was discussed in detail by Rossi \cite{RossiBook}.

Zedler \textit{et al.} \cite{ZedlerPRB09_WeakCoupling} present an interesting analysis of non-Markovian versus Markovian equations in the weak coupling limit
on the example of a quantum dot coupled to contacts with a Lorentzian density of states (i.e. contacts with a finite electron lifetime), thereby going beyond the high bias limit,
and conclude that one must be careful with non-Markovian master equations as they do not necessarily perform better than their Markovian counterparts when non-Markovian effects are strong, and are not in general guaranteed to conserve positivity. The authors compare the exact solution for a single level system with dynamical coarse graining \cite{Schaller08},  non-Markovian master equation, and the Markovian master equation limit \cite{ZedlerPRB09_WeakCoupling}.

\begin{figure}
\includegraphics[width=3 in]{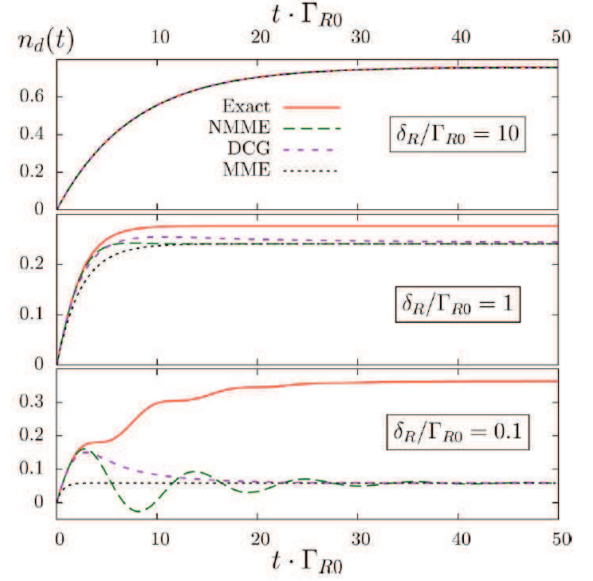}
\caption{Time-dependent occupation probability of the single bound state in a dot coupled to contacts with a Lorentzian density of states with width $\epsilon_R$. Calculation is presented for the exact solution, dynamical coarse graining (DCG), non-Markovian master equation (NMM), and Markovian master equation (MMM). Approximations parameters are $\varepsilon_d=\Gamma_{R,0}$ $\varepsilon_d-\Gamma_{R,0}=\Gamma_L=0.1\Gamma_{R,0}$, where $\varepsilon_d$, $\Gamma_{R,0}$, and $\Gamma_L$ are the dot energy level and the rates of tunneling into the right and left contacts, respectively. Reprinted with permission from \cite{ZedlerPRB09_WeakCoupling},
P. Zedler \textit{et al.}, Phys. Rev. B 80, 045309 (2009). (c) 2009 The American Physical Society
 }\label{fig:VazJPhysConf}
\end{figure}

\subsection{Coarse graining over contact relaxation time}\label{sec:Coarse graining}

As many nanostrucures have no resonances, the work by Novakovic and Knezevic \cite{Knezevic08,NovakovicFdP13} emphasizes the continuous spectrum in the open active region, with forward- and backward-propagating scattering states, whose asymptotic forms are plane waves (a combination of injected and reflected waves in the incoming contact, transmitted wave in the outgoing one). The model interaction Hamiltonian couples each scattering state only with the plane wave with the same wave number $k$ from the injecting contact, i.e.
\begin{eqnarray}\label{eq:Hplus}
H_{\mathrm{int}}=\sum_{k>0}\Delta_k d_k^{\dagger}c_{k,L}+\Delta_{-k} d_{-k}^{\dagger}c_{-k',R}+h.c.
\end{eqnarray}
$c^\dagger_{k,L}$ ($c_{k,L}$) and $c^\dagger_{-k',R}$
($c_{-k',R}$) create (destroy) an electron with a wavevector $k$
in the left and $-k'$ in the right contact, respectively, $d_{k}$ and $d_{k}^\dagger$ do the same for active-region states,
and ${k'}^2-k^2=2m^*eV/\hbar^2$ ($k$ and $k'$ are the wave numbers corresponding to the same energy in the two contacts separated by bias $V$). The hopping
coefficients $\Delta_k$ and $\Delta_{-k}$ are proportional to the current $I_k$ carried by each mode,
$\Delta_k=\frac{I_k}{e\mathcal{T}_k}$, where $\mathcal{T}_k$ is the transmission coefficient of mode $k$ \cite{NovakovicFdP13}.

To obtain a tractable theoretical approach, the full dynamics is coarse grained over the momentum-relaxation time of the contacts. Contact relaxation occurs
on timescales of order $10^1-10^2$ femtoseconds \cite{Osman87,Kriman90}, owing to fast electron-electron scattering that results in a drifted Fermi-Dirac distribution \cite{Lugli85}.

To coarse grain, we partition the time axis into
intervals of length $\tau$, $t_n=n\tau$, so the environment
interacts with the system in approximately the same way during each
interval $[t_n,t_{n+1}]$ \cite{Lidar01},
\begin{equation}\label{eq:difference rho_S}
\frac{d\rho_S}{dt}\approx\frac{\rho_{S,n+1}-\rho_{S,n}}{\tau}=
\mathcal{\overline{K}}_\tau\rho_{S,n},\end{equation} where
$\mathcal{\overline{K}}_\tau=\frac{\int_0^\tau
\mathcal{K}(t')dt'}{\tau}=\frac{\int_{t_n}^{t_{n+1}}\mathcal{K}(t')dt'}{\tau}$
is the averaged value of the map's generator over any interval
$[t_n,t_{n+1}]$ ($\mathcal{K}$ is reset at each $t_n$). If the
coarse-graining time $\tau$ is short enough, then the short-time
expansion of $\mathcal{K}$ can be used to perform the coarse-graining \cite{Knezevic08}.

Each term in the short-time expansion of $\mathcal{K}$ turns out to be a sum of independent contributions over single-particle states,
so in reality we have a multitude of two-level problems, one for each $|k\rangle$,
where the two levels are a particle being in $|k\rangle$ ("+") and a
particle being absent from $|k\rangle$ ("-"). Each such 2-level
problem is cast on its own 4-dimensional Liouville space, with
$\rho_k=\left(\rho^{++}_k,\rho^{+-}_k,\rho^{-+}_k,\rho^{--}_k\right)^\mathrm{T}$
being the reduced statistical operator that describes the occupation of
$|k\rangle$ and evolves according to a master equation
\begin{equation}
\frac{d\rho_k}{dt}=\mathcal{\overline{K}}_{\tau,k}\rho_k .
\end{equation}
The equations for $f_{\pm k}=\rho_{\pm k}^{++}$ become
\begin{subequations}\label{eq:f time evolution}
\begin{eqnarray}
\frac{df_k}{dt}&=&-\tau\Delta_k^2 f_k +\tau \Delta_k^2 f_k^L(k_d (t)),\\
\frac{df_{-k}}{dt}&=&-\tau\Delta_{-k}^2 f_{-k} +\tau \Delta_{-k}^2 f_{-k'}^R (k_d (t)).
\end{eqnarray}
\end{subequations}

\noindent The above equations describe non-Markovian evolution, because drifted Fermi-Dirac distribution functions in the contacts depend on time through the drift wave vector (related to current). As the transient progresses, the current and the charge density in the structure change, which in turn changes the potential profile, the scattering states available to electrons, the transmission coefficients,  and, to a small degree, the interaction matrix elements $\Delta_{\pm k}$, as well as the aforementioned contact distribution functions. Moreover, there may be well-like confined states that cannot be populated by tunneling but only by scattering in the active region. These considerations have been addressed in detail in \cite{NovakovicPhDThesis,NovakovicFdP13}.

\begin{figure}
\includegraphics[width=3 in]{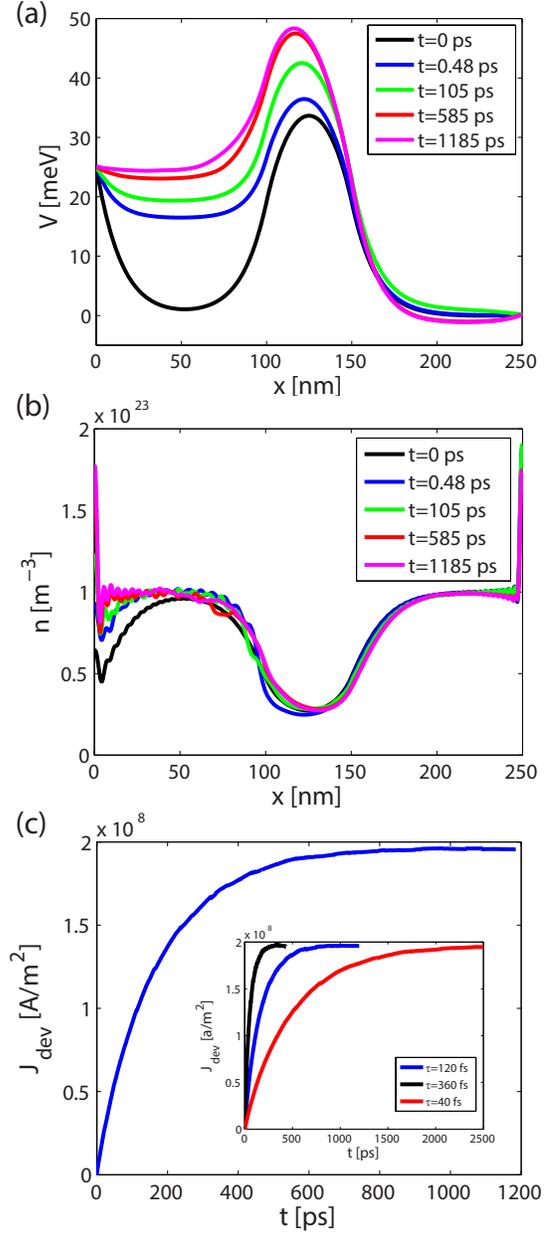}
\caption{Potential (a), charge density (b), and current density (c) in the \textit{nin} diode as a function of time upon the application of -25 mV to the left contact. The \textit{n}-type regions are doped to $10^{17}$ cm$^{-3}$ and contact momentum relaxation time is $\tau$=120 fs, as calculated from the textbook mobility value corresponding to the contact doping density. Inset to panel (c): effect of different contact momentum relaxation times $\tau$ (equal to the coarse-graining times for the active-region dynamics) on the duration of the transient.}\label{fig:Bozidar NIN}
\end{figure}

Figure \ref{fig:Bozidar NIN} depicts the potential, charge density, and current density for a single ellipsoidal valley in an \textit{nin} silicon diode at room temperature. The left and right contacts are doped to $10^{17}$ cm$^{-3}$, whereas the middle region is intrinsic (undoped). In the three main panels, the momentum relaxation time in the contacts is taken to be $\tau=$120 fs, based on the textbook mobility values for the above doping density. The characteristic response time of the current is of order hundreds of picoseconds, three orders of magnitude greater than $\tau$. The transient duration can be thought of as the inverse of a typical $\Delta_k^2\tau$ among the $k$'s participating in the current flow; shorter $\tau$ means weaker coupling and a slower transient (inset).

\section{Conclusion}\label{sec:Conclusion}

Electrons in the active region of nanostructures constitute an open many-body quantum system, coupled with reservoirs of charge, as well as interacting with phonons and photons.
We overviewed the basics of the open system theory, with special focus on the approximations that lead to Markovian and non-Markovian master equations for the reduced statistical operator, and highlighted some recent applications of both types of master equations in quantum transport theory and simulation.

It should be noted that this review did not discuss other widely applied techniques for time-dependent quantum transport, such as the Wigner function simulation \cite{Nedjalkov04,Querlioz2013wigner}, non\-equi\-lib\-rium Green's functions \cite{Jauho94}, time-dependent density functional theory \cite{StefanucciTDDFTPRB04}, Bohmian trajectories \cite{Oriols07}, or full quantum statistics \cite{Blatner00}, which  will receive due attention in other reviews in this special issue. Also, we did not discuss semiconductor Bloch equations \cite{LindbergKochSemiBlochEqns88,Rossi02}, which are often emp\-loy\-ed to address ultrafast optics in semiconductors, and which deserve much more space than available here.

We conclude with some thoughts on the limitations of the master equation framework, as well as potential avenues for further developments.

\textit{Active region/contact partitioning.} An obvious question is where the active region ends and the contacts begin; there is no a good answer to this question, especially for structures that have no resonances. In large and complex physical systems it is impossible to treat all degrees of freedom quantum-mechanically, so a boundary between the quantum and the classical (rapidly dephasing) parts has to be adopted, but a boundary should be moved until convergence is reached and the physics no longer varies with its position \cite{Wang01,Thoss01}.  Rossi \cite{RossiBook} has argued that, in the Wigner function simulations, this seemingly arbitrary introduction of the contact/active region boundary results in artifacts that have conceptual, rather than computational origin.

A related issue is that the reduced statistical operator formalism requires that we be able to write the total many-body Fock space as a tensor product of the Fock spaces of the system and environment, and that we write down an interaction Hamiltonian between the two. With $S$ and $E$ containing electrons, we can try to split the total $S+E$ single-particle Hilbert space into $S$ and $E$ subspaces spanned by specific eigenvectors of the position operator, then construct Fock spaces based on these spatially separated single-particle spaces, and finally form a tensor product of said Fock spaces. Unfortunately, this framework artificially makes the interaction local and is not a good choice for capturing current flow that the full $S+E$ Fock space can describe. Rossi \cite{RossiBook} shows that consistency requires that the effective interaction depend on the overlap between contact states and active region states, where both contact and active region states in principle extend throughout the whole coordinate space.

\textit{Validity of the RWA approximation.} The usual secular or RWA approximation -- assuming
that the system energy levels are so large that the spacing between them is much greater than the system relaxation rate -- works well for optical systems and is amply applied in the derivations of master equations for electronic transport, but may not necessarily hold.  In fact, in nanostructures with a continuum of states, the spacing between relevantly coupled levels is small and easily smaller than the expected system relaxation rate, especially in the case of strong coupling with the contacts. Therefore, the opposite limit, that of quantum Brownian motion \cite{CaldeiraLeggett,Breuer02} may be more applicable in electronic systems with densely spaced system states strongly coupled to the environment. This is a direction in which quantum master equations may have a lot to offer to quantum transport studies \cite{TanimuraJpnPhysCaldeiraLeggettRTD}.

\textit{Uncorrelated initial state.} Considering that, in reality, the contacts and active region share a Fock space, once we partition it into spatially-determined subspaces and if there are no tunnel barriers, it is not easy to justify the approximation of an uncorrelated initial state. Taking a close look into correlated initial states \cite{Velicky_CorrelatedInitCond} can be a very fruitful direction of research, one where a tight coupling between approaches that do not adopt contact/active region partitioning, such as TDDFT, with master equations would likely be necessary.

\textit{High-frequency transport.} Another direction in which the master equation approaches can grow is to look into systems with continua of states and realistic fermionic reservoirs, with a more complete account of intra-reservoir dephasing. This work has opportunities to interface with modern experimental work on GHz-frequency response of nanostructures \cite{GabelliScience06_RC_Circuit,ButtikerNanotech07}.

\textit{Deriving single-particle techniques from statistical operator nonunitary dynamics.} Capturing the entire statistical operator is feasible only in very small systems. With the statistical operator being the ``parent" concept from which single particle quantities such as the density matrix and Green's functions can be derived, it is reasonable to expect that a good non-Markovian approximation for the many-body statistical operator of the electronic system would come first, and from its non-unitary evolution one can further derive single-particle techniques \cite{Knezevic03_1}. An open direction of research is to look at single-particle kinetic approaches that originate from non-Markovian approximations for the evolution of the reduced statistical operator. Time-con\-vo\-lu\-tion\-less non-Markovian equations, thus far underutilized in quantum transport theory, could enable systematic development of single-particle non-Markovian formalisms that are of a fixed order in the interaction.

% BibTeX users please use one of
%\bibliographystyle{spbasic}      % basic style, author-year citations
%\bibliographystyle{spmpsci}      % mathematics and physical sciences
%\bibliographystyle{spphys}       % APS-like style for physics
%\bibliography{}   % name your BibTeX data base

\section{Acknowledgement}
This work has been supported by the U.S. Department of Energy,
Office of Basic Energy Sciences, Division of Materials Sciences
and Engineering, under award DE-SC0008712.

%\bibliography{KnezevicBibliography}

\end{document}